\documentclass[reqno,12pt]{amsart}
\usepackage{amsmath,amsthm,amssymb,amscd}
\usepackage[utf8]{inputenc}
\newcommand{\RR}{\mathbb{R}}
\usepackage{ulem}
\def\beq{\begin{eqnarray}}
\def\eeq{\end{eqnarray}}

\def \mc {\mathcal}

\newcommand{\be}{\begin{equation}}
\newcommand{\ee}{\end{equation}}

\newtheorem{thm}{Theorem}

\begin{document}

\title{Supersymmetry, Holonomy, and Kundt Spacetimes}

\author{J Brannlund}
\address{Department of Mathematics \& Statistics, Dalhousie University,
Halifax, Nova Scotia, Canada B3H 3J5}
\email[J Brannlund]{johanb@mathstat.dal.ca}
\author{A Coley}
% Same address as R. Campbell
\email[A Coley]{aac@mathstat.dal.ca}
\author{S Hervik}
\address[S Hervik]{Faculty of Science and Technology, University of Stavanger, N-4036 Stavanger, Norway}
\email[S.Hervik]{sigbjorn.hervik@uis.no}

%  \author{J Brannlund}
%  \address{ Department of Mathematics \& Statistics, Dalhousie University,
%  Halifax, Nova Scotia, Canada B3H 3J5 }
% \email{johanb@mathstat.dal.ca}
% \author{ A Coley}
% % \address{ Department of Mathematics \& Statistics, Dalhousie University,
% % Halifax, Nova Scotia, Canada B3H 3J5 }
% \email{aac@mathstat.dal.ca}
% \author{S Hervik}
%  \address{Faculty of Science and Technology, University of Stavanger, N-4036 Stavanger, Norway }
%  \email{sigbjorn.hervik@uis.no}

\begin{abstract}
  Supersymmetric solutions of supergravity theories, and consequently
  metrics with special holonomy, have played an important role in the
  development of string theory. We describe how a Lorentzian manifold
  is either completely reducible, and thus essentially known, or not
  completely reducible so that there exists a degenerate holonomy
  invariant lightlike subspace and consequently admits a covariantly
  constant or a recurrent null vector and belongs to the
  higher-dimensional Kundt class of spacetimes. These Kundt spacetimes
  (which contain the vanishing and constant curvature invariant
  spacetimes as special cases) are genuinely Lorentzian and have a
  number of interesting and unusual properties, which may lead to
  novel and fundamental physics.

\end{abstract}

\maketitle

%\newpage

\section{Introduction}

We discuss the geometrical properties of spacetimes in the context of
string theory. In the study of supersymmetry, there are two classes of
solutions. If the spacetime admits a covariantly constant time-like
vector, the spacetime is static and $(1+10)$-decomposable, where the
$10$-dimensional transverse space is Riemannian. The second class of
solutions consists of spacetimes which admit a covariantly constant
light-like vector, and belong to the higher-dimensional Kundt class of
spacetimes.  These spacetimes are genuinely Lorentzian and have many
mathematical properties quite different from their Riemannian
counterparts that can lead to interesting and novel physics. It is
within string theory that the full richness of Lorentzian geometry is
consequently realised, where the Kundt spacetimes play a fundamental
role. Indeed, in gravitational physics the richness of Lorentzian
spacetimes and general relativity is often not fully realised. For
example, in many applications in cosmology Newtonian gravity, or small
deviations thereof, suffices. In many applications of quantum field
theory on a curved background or quantum gravity there exists a unique
time and space and time are essentially treated independently (and the
structure of the Lorentzian manifold is not fully utilized).

In particular, we discuss the work of  \cite{lBaI93,Wu,Leistner},
and show that it follows that a Lorentzian manifold is either
completely reducible, and thus essentially known, or not
completely reducible, which is equivalent to the existence of a
degenerate invariant subspace and entails the existence of a
holonomy invariant lightlike subspace. In this latter case  the
Lorentzian manifold decomposes into irreducible or flat Riemannian
manifolds and a Lorentzian manifold with indecomposable, but
non-irreducible holonomy representation; i.e., with (a
one-dimensional) invariant lightlike subspace. Such Lorentzian
spacetimes admit a covariantly constant or a recurrent null
vector (CCNV/RNV), and contain the vanishing and constant
curvature invariant spacetimes (VSI/CSI) as special cases.

%\newpage

\subsection{Supersymmetry}

Supersymmetric solutions of supergravity theories have played an
important role in the development of string theory. For example,
supersymmetric compactifications may lead to realistic models of
particle physics and a microscopic interpretation of black hole
entropy in string theory is best understood for supersymmetric
black holes. The existence of parallel (Killing) spinor fields, which play a central role in
supersymmetry, accounts for much
of the interest in metrics with special holonomy in mathematical
physics. The existence of a parallel spinor on a Lorentzian manifold
defines a parallel vector field which can be null. Hence the
manifold has an indecomposable, non-irreducible factor. In the
physically important dimensions below twelve the
maximal indecomposable Lorentzian holonomy groups admitting
parallel spinors are known \cite{Bryant,jFF99}.

A systematic classification of supersymmetric $M$-theory vacua
with zero flux (that is, eleven-dimensional Lorentzian manifolds
with vanishing Ricci curvature and admitting covariantly constant
spinors) was provided in  \cite{jFF99}. There are two classes of
solutions. If the spacetime admits a covariantly constant
time-like vector, the spacetimes are static and their
classification reduces  to the classification of $10$-dimensional
Riemannian manifolds with holonomy contained in $SU(5)$ (i.e., to
the classification of Calabi--Yau $5$-folds). The second class of
solutions consists of spacetimes which are not static but which
admit a covariantly constant light-like vector.
Supersymmetric solutions of 11-dimensional supergravity can be
classified according to the holonomy of the supercovariant
derivative arising in the Killing spinor condition. This class can
be extended to M-theory \cite{duff,hull}.

The following results are useful in 11D Supergravity.
{\it  Let $(\mc{M},g_{ab})$ be a Lorentzian manifold. Then there is a
  parallel (i.e., covariantly constant) vector field on $\mc{M}$ if
  and only if the holonomy has a trivial subrepresentation} \cite{besse}.
{\it  Let $(\mc{M},g_{ab})$ be a simply connected, indecomposable,
  reducible Lorentzian manifold with abelian holonomy algebra
  $\mathbb{R}^{m-2}$. Then the manifold admits parallel
  spinors whose corresponding vectors are lightlike} \cite{Leistner}.
The classification can be further refined
by the holonomy group of the spacetime.  In the Lorentzian case
there are subgroups of the Lorentz group which act reducibly yet
indecomposably on the spacetime that play an important role. We shall
discuss these results in more detail later.

Summarizing the situation in eleven-dimensional supergravity \cite{jFF99}:

\subsubsection{Static vacua}

If the spacetime admits a covariantly constant time-like vector, then
the holonomy group must be contained in the subgroup $Spin(10) \subset
Spin(10,1)$, leaving a time-like vector invariant.  Spacetimes with
holonomy group $H\subset SU(5)$ always admit a Ricci-flat metric and
hence, equipped with this metric, they satisfy the supergravity
equations of motion. Such spacetimes contain a time-like Killing
vector and are static and consequently
locally isometric to a product $\RR \times X$ with metric
\begin{equation*}
  ds^2 = - dt^2 + ds^2(X)~,
\end{equation*}
where $X$ is any Riemannian $10$-manifold with holonomy contained
in $SU(5)$; that is, a Calabi--Yau $5$-fold.

\subsubsection{Non-static vacua}

If the spacetime admits a covariantly constant null vector, the
isotropy subgroup of a null spinor is contained in the isotropy
subgroup of the null vector, which in arbitrary dimensions is
isomorphic to the spin cover of
\begin{equation*}
 ISO(n-2) = SO(n-2) \ltimes \RR^{n-2} \subset SO(n-1,1)~.
\end{equation*}
For $n\leq 5$ this means the holonomy group is $\RR^{n-2}$,
which implies that
the metric is Ricci-null. Bryant \cite{Bryant} has recently
written down the most general local metric with this holonomy
(particularly for a eleven-dimensional spacetime).

\subsection{Recurrent Null Vector}

Another group of interest in Lorentzian signature in $n$
dimensions is the maximal proper subgroup of the Lorentz group,
$\rm Sim(n-2)$. The Einstein metric of $\rm Sim(n-2)$
holonomy, with and without a cosmological constant, was studied in
\cite{GibbonsPope}.

In general, time-dependent solutions cannot be supersymmetric.
However, a time-dependent solution can arise from the dimensional
reduction of a time-independent solution in one higher dimension
that admits a Killing spinor. Because the Killing spinor is not
boost invariant, it does not descend to the lower-dimensional
spacetime which is not supersymmetric.

Multi-centre metrics can arise by dimensional reduction from a
metric with a reduced holonomy group. An
example is when the Killing vector field is (covariantly constant
and) null.  The holonomy is then reduced to the abelian subgroup
${\Bbb R}^{n-2}$ of the $n$-dimensional Lorentz group $SO(n-1,1)$.
These Brinkmann waves \cite{Brinkmann} (or CCNV spacetimes)
necessarily have vanishing Ricci scalar curvature. They can be
used to obtain time-independent and time-dependent extremal
Kaluza-Klein multi black holes and  multi Kaluza-Klein monopole
metrics.

Dimensional reduction on a null Killing vector that is not
necessarily covariantly constant has also been studied. Indeed,
solutions with  non-vanishing spatial curvature from Einstein
metrics in one higher dimension with reduced holonomy and with
non-vanishing cosmological constant must necessarily be obtained for a
holonomy group which is larger than that of ${\Bbb R} ^{n-2}$.

The ${\rm Sim}(n-2)$ holonomy metrics
are of interest in M-theory and string theory. In any of the
metrics of $\rm Sim(n-2)$ holonomy there exists a preferred spinor
which generalises the covariantly-constant spinor that gives rise
to supersymmetric CCNV metrics; however, this preferred spinor is not
associated with any supersymmetry.

%\newpage

\section{Holonomy Theorem}

In the Riemannian case, the de Rham  decomposition theorem
\cite{dR} states that if the {\it { holonomy group}} (see Appendix) of a
simply-connected Riemannian manifold $M$ at $m \in M$, $\Phi$,
preserves a proper subspace of the tangent space (i.e., is
reducible), then the  tangent space is decomposable (into holonomy
invariant subspaces)  and M is (locally) isometric to a product
manifold. That is, a Riemannian manifold  is locally a product of
Riemannian manifolds with irreducible  holonomy algebras. This
allows one to restrict oneself to groups acting irreducibly, and
leads to the classification of
Riemannian holonomy groups of Berger \cite{Berger}.

The possible restricted holonomy groups of irreducible Riemannian
manifolds are known~\cite{Berger,Bryant}.  In the familiar
positive-definite case, Berger's classification \cite{Berger} implies
that a simply connected, irreducible, non-symmetric manifold can be
Calabi-Yau, with $SU(n)$ holonomy in $2n$ dimensions, hyper-K\"ahler,
with $Sp(n)$ holonomy in $4n$ dimensions, have $G_2$ holonomy in
seven-dimensions and Spin(7) holonomy in eight dimensions. In each
case, there exists a metric with the given holonomy and vanishing Ricci tensor. An
exceptional case with $Sp(n)Sp(1)$ holonomy and Einstein metric also
exists.

\subsection{Lorentz Manifolds}\label{ssec:psdoRgen}

The {\it {Ambrose-Singer theorem}} \cite{KN} says that the Lie
algebra of the holonomy group of the manifold is determined by the
curvature of the corresponding connection; thus the existence of a
parallel tensor constrains the holonomy algebra and determines the
product structure (a precise statement, and a number of
appropriate definitions, are given in the Appendix).

The classification of holonomy groups in Lorentzian spacetimes
is quite different since the Riemannian decomposition theorem does not apply
without modification \cite{lBaI93}. It is
important to extend the de Rham decomposition theorem to the
Lorentzian case \cite{KN,Wu,dR}.

The classification of the holonomy-irreducible case proceeds much
as in the positive definite case. In the Lorentzian  case, there
are spacetimes such that no proper subgroup of the Lorentz group
acts irreducibly. A general survey of the pseudo-Riemannian
case when the holonomy acts irreducibly, particularly regarding
the existence of parallel spinor fields, was given in~\cite{hBiK99}. It is
this difference that makes classifying the possible
pseudo-Riemannian metrics having parallel spinors more
difficult. Thus, the Lorentzian analogue of the decomposition
theorem, due to Wu \cite{Wu}, requires the weaker notion of weak
irreducibility.

Wu's theorem \cite{Wu} asserts that {\it every simply-connected,
complete semi-Riemannian manifold is isometric to a product of
simply-connected, complete {semi-Riemannian} manifolds, of which
one can be flat and all others are indecomposable or
``weakly-irreducible''} (i.e., with no non-degenerate invariant
subspace under holonomy representation). For a Riemannian manifold
this theorem asserts that the  holonomy representation is
completely reducible; i.e., decomposes into factors which are
trivial or irreducible, and are again Riemannian holonomy
representations.  For pseudo-Riemannian manifolds
indecomposability is not the same as irreducibility. We can have
degenerate invariant subspaces under holonomy representation.

This exotic class of holonomy groups, acting
reducibly but indecomposably, are of particular interest in
supersymmetry. The classification of these spacetimes is still not
complete \cite{jFF99} (see also \cite{aI96,lBaI93,galaev}).
All simply connected irreducible
non-locally symmetric Lorentzian manifolds admitting parallel
spinors were studied by \cite{aI96}.
All irreducible factors are known by the Berger classification of
possible irreducible semi-Riemannian holonomy groups
\cite{Berger}. Bryant determined the local generality of pseudo-Riemannian
metrics with parallel spinors, with and without imposing the
Ricci-flat condition  \cite{Bryant}.

\subsection{Classification}

Thus in order to classify holonomy groups of simply-connected
Lorentzian manifolds it is necessary to find the possible holonomy
groups of indecomposable (i.e., weakly-irreducible), but
non-irreducible Lorentzian manifolds.  The holonomy algebra of
such a manifold of dimension $n$ is contained in the
$\textrm{Sim}(n-2)$ algebra.  The projections of this holonomy
algebra, classified into four types based on the possible
projections on $\mathbb{R}$ and $\mathbb{R}^{(n-2)}$, were studied
in \cite{lBaI93}. A decomposition property for the
$\mathfrak{so}(n-2)$--projection was also found in \cite{lBaI93}
(i.e., there is a decomposition of the representation space into
irreducible components and of the Lie algebra into ideals which
act irreducibly on the components).

Thus, it suffices to study irreducibly acting groups or algebras, a
fact which is necessary for obtaining a
classification.
An algebraic criterion on the $\mathfrak{so}(n-2)$--component of
an indecomposable, reducible, simply-connected Lorentzian
manifold, in analogy to the well known Berger criterion
for holonomy algebras, was derived by Leistner \cite{Leistner}.
Furthermore, it was shown that every irreducible weak-Berger
algebra which is contained in $\mathfrak{u}((n-2)/2)$ is a Berger
algebra, and from the decomposition property  is, in particular, a
Riemannian holonomy algebra.  In addition, the
result was then repeated for a simple weak-Berger algebra which is
not contained in $\mathfrak{u}((n-2)/2)$ \cite{Leistner}. Finally,
it was shown that there are no semisimple, non simple, irreducibly
acting Lie algebras not contained in $\mathfrak{u}((n-2)/2)$
which are weak-Berger but not Berger \cite{galaev}  (the converse
results were also discussed).

It therefore follows that
{\it every $SO(n-2)$--projection of an indecomposable,
non-irreducible Lorentzian holonomy group is a Riemannian holonomy
group \cite{Leistner}}.

The holonomy group at a point $p$ in an indecomposable
non-irreducible Lorentzian manifold, acting on $T_p M$, then has a
null, one-dimensional invariant subspace, which is equivalent
to the existence of a recurrent null vector field.
The existence of the recurrent vector leads to the holonomy algebra
$(\mathbb{R} \oplus \mathfrak{so}(n-2)) \ltimes \mathbb{R}^{n-2}$, which
is the $\textrm{Sim}(n-2)$ algebra. The corresponding group is
the maximal proper subgroup of the Lorentz group $SO(n-1,1)$, a
subgroup of dimension $(n^2-3n+4)/2$. ${\rm Sim} (n-2)$ holonomy
implies the existence of a recurrent null vector \cite{GibbonsPope}.

With
respect to the four projections of the holonomy algebra on the
$\mathbb{R}$-- and on the $ \mathbb{R}^{(n-2)}$--components, we have the following
\cite{lBaI93}: For the types $I$ and $II$ the holonomy is equal to
$(\mathbb{R} \oplus \mathfrak{g}) \ltimes \mathbb{R}^{(n-2)}$ and
 $\mathfrak{g}\ltimes\mathbb{R}^{(n-2)}$, respectively. In the case of types
$II$ and $IV$ the projection on $\mathbb{R}$ is zero, which
implies the existence not only of a recurrent null vector
field but also of a parallel one. In the case of types $III$ and $IV$
the $\mathbb{R}$--  the $\mathbb{R}^{(n-2)}$--
components are coupled to the $\mathfrak{so}(n-2)$--component.

All candidates to the weakly-irreducible not
irreducible holonomy algebras of Lorentzian manifolds are known.
To complete the classification of holonomy algebras it was proved that
all Berger algebras can be realised as holonomy algebras of
Lorentzian manifolds \cite{galaev}.

\subsubsection{Summary}

As a result, we have the following  \cite{lBaI93,Wu,Leistner}:

{\it For a  Lorentzian manifold, $M$, the de Rham/Wu--decomposition
yields the following two cases. 1. {Completely reducible:} Here
$M$ decomposes  into irreducible or flat Riemannian manifolds
and a manifold which is an irreducible or flat Lorentzian manifold
or $(\mathbb{R},-dt)$. The irreducible Riemannian holonomies are
known, as well as the irreducible Lorentzian one, which has to be
the whole of $SO(1,n-1)$. 2. {Not completely reducible:} This is
equivalent to the existence of a degenerate invariant subspace and
entails the existence of a holonomy invariant lightlike
subspace. The Lorentzian manifold decomposes into irreducible or
flat Riemannian manifolds and a Lorentzian manifold with
indecomposable, but non-irreducible holonomy representation; i.e.,
with (a one-dimensional) invariant lightlike subspace.}

\subsection{The 4D case}

In four dimensions (4D) there are 15 holonomy subgroups,
labelled $R_1$ -- $R_{15}$ \cite{HallBook}. $R_1$ is the trivial group
and $R_5$ cannot exist as a spacetime holonomy. $R_{15}$ is the full
Lorentz group and corresponds to the irreducible case. The
non-degenerately reducible subgroups are $R_2$, $R_3$, $R_4$, $R_6$,
$R_7$, $R_{10}$ and $R_{13}$ (the corresponding spacetimes include the
$1+3$ and $2+2$ decomposable spacetimes; e.g., $R_{13}$). Finally, the
reducible but degenerate subgroups are $R_8$, $R_9$, $R_{11}$,
$R_{12}$ and $R_{14}$; spacetimes with $R_8$ and $R_{11}$ holonomy
admit a CCNV, while spacetimes with $R_9$, $R_{12}$ and $R_{14}$
holonomy admit a RNV.

The holonomy group in a 4D Lorentzian spacetime
was reviewed by Hall \cite{HallBook}; the holonomy algebras and
associated RNV and CCNV in 4D were given explicitly
in the table therein. The holonomy group can be used to
find {\it all} metrics on $M$ \cite{Hall1988}.
 The well-known classification of
pseudo-Riemannian metrics with parallel spinors in 4D
has been reviewed by Bryant \cite{Bryant};
the Lorentzian case was studied in \cite{Tod}.

%\newpage

\section{Kundt spacetimes}

\subsection{Higher dimensional Kundt spacetimes:}

The n-dimensional {\it{Kundt metric}}  can be written \cite{Coley}
 \begin{equation} d s^2=2d u\left[d v+H(v,u,x^k)d u+W_{i}(v,u,x^k)d x^i\right]+
 g_{ij}(u,x^k)d x^id x^j. \label{Kundt}\end{equation} The~metric functions
 $H$, $W_{i}$ and $g_{ij}$ satisfy the Einstein equations ($i,j = 2, ..., N-2$).
The metric (\ref{Kundt}) possesses a null vector field
${\ell} = \frac{\partial}{\partial{v}}$ which is  geodesic,
non-expanding, shear-free and non-twisting (i.e., the
Ricci rotation coefficients $L_{ij} \equiv
\ell_{i;j}=0$).
Kundt spacetimes
are of Riemann type II (or simpler) \cite{Coley}.

\subsubsection{Covariantly constant null vector (CCNV)}

In general, the generalized Kundt metric  has the non-vanishing
Ricci rotation coefficients $L$ and $L_{i}$. From  $L =0$ we
obtain $H_{,v} =0$ and $L_i = 0$ implies $W_{i,v} = 0$
\cite{CCNV}. The remaining transformations can be used to
further simplify the remaining non-trivial metric functions.
Thus, the aligned, repeated, null vector $\ell$ of (\ref{Kundt}) is a
null Killing vector (KV) if and only if $H_{v}=0$ and
$W_{i,v}=0$ (whence the
 metric no longer has any $v$
dependence).  Furthermore, it follows
that if $\ell$ is a null KV then it
is also covariantly constant.  Without any further restrictions, the higher
dimensional metrics
admitting a null KV are of Ricci and Weyl type {II}.
Therefore, the most general metric that admits a covariantly constant null vector
(CCNV) is (\ref{Kundt})
with $H = H(u,x^k)$ and $W_{i} = W_{i}(u,x^k)$
\cite{CCNV}; we shall refer to this as a {\it a Kundt-CCNV} metric.

\subsubsection{Recurrent null vector (RNV)}

As shown in \cite{GibbonsPope}, the null vector field of a metric
with ${\rm Sim}(n-2)$ holonomy is recurrent, and it follows  that
the null congruence is geodesic,  twist-free, expansion-free and
shear-free; i.e., the metrics belong to the class of Kundt metrics
\cite{Coley}.  Metrics with ${\rm Sim}(n-2)$ holonomy may be cast
in the form of (\ref{Kundt})  with $W_{i} = W_{i}(u,x^k)$ (i.e.,
independent of $v$); called {\it a Kundt-RNV} metric or a metric
of Walker form. For the {\rm Sim}$(n-2)$ metrics there are no
further restrictions on the metric functions.

\subsubsection{Constant curvature invariants (CSI)}

If we require that all of the curvature
 invariants are constants, it follows that the metric function $H(v,u,x^k)$ in the Kundt metric
(\ref{Kundt}) only  contains terms polynomial and second order in $v$; i.e.,
\[ H(v,u,x^k)=v^2 \tilde \sigma +vH^{(1)}(u,x^k) +H^{(0)}(u,x^k),\]
where  $\tilde \sigma = (4\sigma+W^{(1)i}W^{(1)}_i)/8$ and 
$\sigma $ is a constant.
In addition, for a Kundt-CSI spacetime there exists a ($u$-dependent) diffeomorphism
such that the transverse metric can be made $u$-independent. Furthermore, the transverse
metric $d{\tilde s}^2$ is locally homogeneous.
Hence, there is no loss of generality in assuming that $d{\tilde s}^2$
is a locally homogeneous space, $M_{\rm Hom}$.

\subsubsection{Vanishing curvature invariants (VSI)}
In this case all curvature invariants to all orders vanish, which implies that we can set
\[ \sigma=0, \quad g_{ij} dx^i dx^j=\delta_{ij}dx^i dx^j.\]
(This is the $W^{(1)}_i=0$ case of the general VSI metrics.) The Riemann tensor is of type
III (or simpler).

\subsection{Supergravity theories}

The VSI and CSI spacetimes are of particular interest
since they are solutions of supergravity or superstring theory, when supported by
appropriate bosonic fields.
The supersymmetry properties of these
spacetimes have also been discussed.

In \cite{VSISUG}
it was shown that the higher-dimensional
VSI spacetimes with fluxes and dilaton are solutions of type IIB
supergravity.
Exact VSI solutions of IIB supergravity with NS-NS and
RR fluxes and dilaton have been constructed. The solutions are classified according to their
Ricci type (N or III). The Ricci type N solutions are
generalizations of pp-wave type IIB supergravity solutions. The
Ricci type III solutions are characterized by a non-constant
dilaton field. In particular, it was shown
that all Ricci type N VSI spacetimes are solutions of supergravity
(and that Ricci type III VSI spacetimes are also
supergravity solutions if supported by appropriate sources), and similar
results are expected in all supergravity theories. It was also argued
that, in general, the VSI spacetimes are exact string solutions to all orders
in the string tension. It is known that the higher-dimensional  pp-wave
spacetimes are exact solutions in string theory, in type-IIB superstrings with an
~R-R five-form, and also with NS-NS form fields \cite{ss1,ss2}.

A number of $CSI$ spacetimes are also known to be
solutions of supergravity theory when supported by
appropriate bosonic fields \cite{CFH}.
It is known that $AdS_d \times S^{(N-d)}$ (in short $AdS\times S$)
is an exact solution of supergravity (and preserves the maximal
number of supersymmetries) for certain values of (N,d) and for
particular ratios of the radii of curvature of the two space
forms. Such spacetimes  (with $d=5,N=10$) are supersymmetric
solutions of IIB supergravity (and there are analogous solutions
in $D=11$ supergravity).  $AdS \times S$ is an
example of a CSI spacetime \cite{CSI}. There are a number of other
CSI spacetimes known to be solutions of supergravity and admit
supersymmetries; namely, generalizations of $AdS \times S$ and (generalizations of) the chiral
null models \cite{hortseyt}. The  Weyl type $III$ $AdS$ gyraton \cite{FZ} (which is a CSI
spacetime with the same curvature invariants as pure AdS)
is a solution of gauged supergravity
(the $AdS$ gyraton can be cast in the Kundt form \cite{CFH}).

Some explicit examples of CSI supergravity
spacetimes were constructed in \cite{CFH}
by generalising
a homogeneous Einstein spacetime,
$(\mathcal{M}_{\text{Hom}},\tilde{g})$, of Kundt form
to an inhomogeneous
spacetime, $(\mathcal{M},{g})$, by including arbitrary functions
${W}_{ i}^{(0)}(u,x^k)$, ${H}^{(1)}(u,x^k)$ and
${H}^{(0)}(u,x^k)$.
In the examples $(\mathcal{M}_{\text{Hom}},\tilde{g})$ was chosen to be a
regular Lorentzian Einstein solvmanifold or the
transverse space was chosen to be the Heisenberg group, $SL(2,\mathbb{R})$,
or the 2-sphere, $S^2$.

The supersymmetry properties of VSI
spacetimes have also been studied. It is known that in general if a spacetime admits a
Killing spinor, it necessarily admits a null or timelike Killing
vector. Therefore, a necessary (but not sufficient) condition for
a particular supergravity solution to preserve some supersymmetry
is that the spacetime possess such a Killing vector.
Therefore the supersymmetry properties of VSI type IIB
supergravity solutions with a CCNV were studied \cite{VSISUG}.
Supersymmetry was also studied in the CSI-CCNV
subclass of supergravity spacetimes  \cite{CFH}.

%\newpage

\section{Discussion}

A Lorentzian manifold admitting an indecomposable but
non-irreducible holonomy representation (i.e., with a
one-dimensional invariant lightlike subspace) is a  $CCNV$ or
$RNV$ (Kundt) spacetime, which contains the $VSI$ and $CSI$
subclasses as special cases.
Therefore, the  Kundt spacetimes that are of particular physical
interest are degenerately reducible, which leads to complicated
holonomy structure and various degenerate mathematical properties.
Such spacetimes have a number of other interesting and unusual
properties, which may lead to novel and fundamental physics.
Indeed, a complete understanding of string theory is not possible
without a comprehensive knowledge of the properties of the Kundt
spacetimes. 

For example, in general a Lorentzian spacetime is
completely classified by its set of scalar polynomial curvature
invariants. However, this is not true for Kundt spacetimes
\cite{CSI2} (i.e., they have important geometrical information
that is not contained in the scalar invariants). All {$VSI$}
spacetimes and {$CSI$} spacetimes that are not locally homogeneous
(including the important $CCNV$ and $RNV$ subcases) belong to the
Kundt class \cite{Coley}. In these spacetimes all of the scalar
invariants are constant or zero. This leads to interesting
problems with any physical property that depends essentially on
scalar invariants, and may lead to ambiguities and pathologies in
models of quantum gravity or string theory.

As an illustration, in many theories of fundamental physics there
are geometric classical corrections to general relativity.
Different polynomial curvature invariants (constructed from the
Riemann tensor and its covariant derivatives) are required to
compute different loop-orders of renormalization of the
Einstein-Hilbert action \cite{Dixon}. In specific quantum models
such as supergravity there are particular allowed local
counterterms (such as, e.g., the square of the quadratic
Bel-Robertson tensor at 3-loop order in D=4, N=1 SUGRA, 2-loop
divergences involving terms quartic in curvature in D=11 SUGRA,
etc. \cite{sugra}).

In particular, a classical solution is called {\it universal}  if
the quantum correction is a multiple of the metric. In \cite{CGHP}
metrics of holonomy $\mathrm{Sim}(n-2)$ were investigated, and it
was found that all 4-dimensional $\mathrm{Sim}(2)$ metrics are
universal and consequently can be interpreted as metrics with
vanishing quantum corrections and are automatically solutions to
the quantum theory. The $RNV$ and $CCNV$ (Kundt) spacetimes
therefore play an important role in the quantum theory, regardless
of what the exact form of this theory might be.

Finally, in the domain of Planck scale curvatures, the character
of gravity may change radically due to its underlying quantum
nature. The expectation is that singularities will be ``resolved''
in the correct theory of quantum gravity. Indeed, spacetimes which
are singular in general relativity can be completely nonsingular
in string theory \cite{Horo}. However, it is not true that all
singularities are removed in string theory. The $VSI$ and $CSI$
Kundt spacetimes with arbitrary $u$ dependence are exact solutions
to string theory \cite{ss1}.  However, if any of the metric
functions diverge as $u \rightarrow u_0$, then the Kundt spacetime
is singular. By studying the string propagation in this
background, the string does not always have well behaved
propagation through this singularity since the divergent tidal
forces cause the string to become infinitely excited \cite{ss1}.
Indeed, it has been argued that on physical grounds,  any
reasonable theory will not ``resolve'' certain classes of timelike
singularities, since the elimination of these singularities would
lead to a theory without a stable ground state \cite{Hormy}.

%\newpage

\section{Appendix}

\subsubsection{Definitions}

The {\it {holonomy group}} of $M$ is denoted by $ \Phi$. If one
restricts to members which are continuously or smoothly homotopic to
zero one arrives at the {\it {restricted holonomy group}} of $M$
denoted abstractly by $ \Phi ^0$. In fact $ \Phi ^0$ is the identity
component Lie subgroup of $\Phi$. If $M$ is simply connected $ \Phi ^0
= \Phi$. The associated holonomy algebra is denoted $ \phi$.

The {\it {infinitesimal holonomy algebra}} of $M$ at $p$, $ \phi
'_p$, is a subalgebra of $\phi$ for each $ p \in M$. By
definition, $ \phi '_p$ is the Lie algebra spanned by $R(X,Y)$,
$\nabla R (X,Y,Z)$, $\nabla \nabla R(X,Y,Z,W)$, ... If the
infinitesimal holonomy algebra is determined by the Riemann tensor
alone, the manifold is called {\it
 perfect}. The corresponding unique connected Lie subgroup arising from $ \phi
'_p$ is referred to as the {\it {infinitesimal holonomy group}} of
$M$ at $p$ and is denoted by $ \Phi '_p$ . Under weak assumptions
on the manifold, $\Phi^0=\Phi'$\cite{ozeki}.

The holonomy group $\Phi$ of $M$ is called {\it {reducible}} if
for some (and hence any) $ p \in M$ and for some non-trivial
proper subspace $ V \subseteq T_p M$, $V$ is invariant under each
member of $ \Phi _p$ . Otherwise $\Phi$ is called {\it {
irreducible}}. Such a subspace $V$ is called {\it {holonomy
invariant}}. Further, $\Phi$ is called {\it {non-degenerately
reducible}} if a (non-trivial proper) non-null holonomy invariant
subspace of $ T_p M$ exists at some (and hence every) $ p \in M$.
In such a situation  $ T_p M$ is said to be {\it{decomposable}},
{\it{weakly reducible} } or {\it{non-degenerately reducible}}. If
one instead considers the restricted holonomy group $\Phi ^0$, one
speaks of {\it{strictly irreducible}} and {\it{strictly
holonomy-indecomposable}} manifolds.

If $\Phi$ is reducible, but not non-degenerately reducible, it is
called {\it {degenerately reducible}}. Therefore, in the
Lorentzian case, there is a distinction to be made between a
metric being holonomy-irreducible (no parallel subbundles of the
tangent bundle), being holonomy-indecomposable (no parallel
splitting of the tangent bundle), and being indecomposable (no
local product decomposition of the metric)~\cite{Wu}.

\subsubsection{The Ambrose-Singer theorem}

The theorem \cite{KN} is as follows:

\begin{thm}
  (Ambrose-Singer) The Lie algebra $\phi_p$ of $\Phi_p$ (and of
  $\Phi$) is generated by

  \begin{displaymath}
    R_{abcd}X_q^c Y_q^d
  \end{displaymath}
  where $X_q$ and $Y_q$ range over all possible tangent vectors at $q$
  and the point $q$ ranges over all points that can be joined to $p$
  by a parallel-transported curve.
\end{thm}

%\newpage

\section*{Acknowledgment}

This work was supported, in part, by NSERC. 

%%We wish to thank....

\providecommand{\bysame}{\leavevmode\hbox to3em{\hrulefill}\thinspace}


\begin{thebibliography}{10}


\bibitem{lBaI93}  L. B\'erard-Bergery and A. Ikemakhen,
Proc. Sympos. Pure Math. (AMS), \textbf{54} 27 (1993).

\bibitem{Wu}   H. Wu,  Pacific J. Math.
 \textbf{20}  351 (1967) \& Illinois J. Math.
  \textbf{8}  291 (1964).



\bibitem{Leistner} T. Leistner,  arXiv:math.dg/0305139 \& 
arXiv:math.dg/0309274

\bibitem{Bryant} R.L.~Bryant, Ann.  Math.
\textbf{126}  525 (1987) \&  arXiv:math.dg/0004073



\bibitem{jFF99}
J. Figueroa-O'Farrill, Class. Quant. Grav. {\bf 17} 2925 (2000).


\bibitem{duff}
M. J. Duff and J. T. Liu,  arXiv:hep-th/0303140.
%" Hidden spacetime symmetries and generalized holonomy in M-theory"

\bibitem{hull}
C.~Hull, arXiv:hep-th/0305039.
%``Holonomy and symmetry in M-theory,''


\bibitem{besse}  A. L. Besse, {\it Einstein Manifolds} (Springer Verlag, 1987).



\bibitem{GibbonsPope} G. W.  Gibbons and C. N.  Pope, arXiv:0709.2440.



\bibitem{Brinkmann}
H.W.~Brinkmann,  Math. Ann. \textbf{94} 119 (1925).

\bibitem{dR}
G.~De Rham, Comm. Math.
  Helv. \textbf{26} 328 (1952).


\bibitem{Berger}
M. M. Berger, Bull. Soc. Math. France \textbf{83} 279 (1955) \&
Ann. Sci. Ecole Norm. Sup. {\bf 74} 85 (1975).


\bibitem{KN} S. Kobayashi and K.
Nomizu, {\it Foundations of Differential Geometry and its
Applications: Vol.1} (Interscience, New York 1963).




\bibitem{hBiK99}
 H. Baum and I. Kath,
Ann. Glob. Anal. Geom. \textbf{17}  1 (1999).


\bibitem{aI96}
A. Ikemakhen, Ann. Sci. Math. Qu{\'e}bec, {\bf 20}  53 (1996).

\bibitem{galaev}
A. S. Galaev, arXiv:math.dg/0304407 \& arXiv:math.dg/0502575
%\newblock {The spaces of curvature tensors for holonomy algebras of Lorentzianmanifolds}, 2003. arXiv:math.DG/0304407;metrics that realize all Lorentzian holonomy algebras. arXiv:math.DG/0502575.



\bibitem{HallBook} G. S. Hall, {\it Symmetries and curvature structure in general Relativity} (World
Science, Singapore, 2004).



\bibitem{Hall1988}  G. S.  Hall, Gen. Rel. Grav. {\bf20}
399 (1988).


\bibitem{Tod} K.~P.~Tod,   Phys.\ Lett.\ B {\bf 121}  241 (1983) \&
 Class.\ Quant.\ Grav.\  {\bf 12} 1801 (1995).




\bibitem{Coley} A. Coley,
Class. Quant. Grav. {\bf 25} 033001 (2008).





\bibitem{CCNV} T. Ortin, {\it Gravity and Strings} (Cambridge
University Press, 2000); A. Coley, A. Fuster, S. Hervik and N. Pelavas, Class. Quant. Grav. {\bf 23},
 7431 (2006). A. Coley, N. Pelavas and D. McNutt,  preprint (2008).


\bibitem{VSISUG} A. Coley, A. Fuster, S. Hervik and N. Pelavas, JHEP {\bf 0705}, 032 (2007).



\bibitem{ss1}
D. Amati and C. Klim\v c\'\i k, Phys. Lett. B {\bf 219}, 443
(1989); G.T. Horowitz and A.R. Steif, Phys. Rev. Lett. {\bf 64}
260 (1990); A.A. Coley, Phys. Rev. Lett. {\bf 89}, 281601 (2002).


\bibitem{ss2}
R.~R.~Metsaev and A.~A.~Tseytlin, Phys. Rev.  D {\bf 65}, 126004
(2002); M. Blau et al., JHEP {\bf 0201}, 047 (2002); P. Meessen,
Phys. Rev. D {\bf 65}, 087501 (2002); J. G. Russo and A.A.
Tseytlin, JHEP {\bf 0209}, 035 (2002); J.~Maldacena and L.~Maoz,
JHEP {\bf 0212}, 046 (2002).



\bibitem{CFH} A. Coley, A. Fuster and S. Hervik, arXiv:0707.0957


\bibitem{CSI}    A. Coley, S. Hervik and N.
Pelavas, Class. Quant. Grav. {\bf 23}, 3053 (2006).

\bibitem{hortseyt} G. T. Horowitz  and A. A. Tseytlin, Phys. Rev. D {\bf 51}, 2896
(1995).

\bibitem{FZ} V. P. Frolov and A. Zelnikov, Phys. Rev. D {\bf 72},
104005 (2005); V. P. Frolov and D. V. Fursaev, Phys. Rev. D {\bf
71}, 104034 (2005).

\bibitem{CSI2} A. Coley, S. Hervik and N.
Pelavas, Class. Quant. Grav. {\bf 25},  025008 (2008).


\bibitem{Dixon} L. Dixon, J. Harvey, C. Vafa, and E. Witten, J.
Nucl. Phys., B261, 678 (1985); M. Rocek and E. Verlinde, J. Nucl.
Phys., B373, 630 (1992).

\bibitem{sugra} S. Deser, J. Kay and K. Stelle, Phys. Rev. Letts. {\bf 38}, 527
(1977); Z. Bern et al., Nucl. Phys. {\bf B530} 401 (1998); S.
Deser and D. Seminara, Phys. Rev D {\bf 62} 084010 (2000).


\bibitem{CGHP}
A.A. Coley, G.W. Gibbons, S. Hervik and C.N. Pope, arXiv:0803.2438
%Metrics With Vanishing Quantum Corrections,



\bibitem{Horo} G. T. Horowitz, New J. Phys. {\bf 7} 201 (2005).

\bibitem{Hormy} G. T. Horowitz and R. Myers, Gen. Rel. Grav. {\bf 27} 915
(1995).


\bibitem{ozeki} H. Ozeki,  Nagoya Math. J. {\bf 10} 105 (1956).




\end{thebibliography}
\end{document}